\def\year{2019}\relax
\documentclass[letterpaper]{article} %DO NOT CHANGE THIS
\usepackage{aaai19}  %Required
\usepackage{times}  %Required
\usepackage{helvet}  %Required
\newcommand{\eg}{\emph{e.g. }}

\usepackage{courier}  %Required
\usepackage{url}  %Required
\usepackage{graphicx}  %Required
\usepackage{amsmath}

\makeatletter 
\newcommand{\Rmnum}[1]{\expandafter\@slowromancap\romannumeral #1@}
\makeatother

\usepackage{wrapfig}
\usepackage{float}
\usepackage{subfig}
\newcommand{\ssection}[1]{\noindent {\bf #1} }

\begin{document}
% The file aaai.sty is the style file for AAAI Press 
% proceedings, working notes, and technical reports.
%
\title{3D Face Synthesis Driven by Personality Impression}
\author{Yining Lang$^{1}$ \quad Wei Liang$^{1}$ \quad Yujia Wang$^{1}$ \quad Lap-Fai Yu$^{2}$ \\
$^1${Beijing Institute of Technology} \quad  $^2${University of Massachusetts Boston} \\
{\tt\small \{langyining,liangwei,wangyujia\}@bit.edu.cn  craigyu@cs.umb.edu}}
%Association for the Advancement of Artificial Intelligence\\
%2275 East Bayshore Road, Suite 160\\
%Palo Alto, California 94303\\
%}

\maketitle
\begin{abstract}
Synthesizing 3D faces that give certain personality
impressions is commonly needed in computer games, animations,  and
virtual world applications for producing realistic virtual
characters.
 In this paper, we propose a novel approach to synthesize 3D faces based on
  personality impression for creating virtual characters. Our approach
  consists of two major steps. In the first step, we train classifiers
  using deep convolutional neural networks on a dataset of images with
  personality impression annotations, which are capable of predicting
  the personality impression of a face. In the second step, given a 3D
  face and a desired personality impression type as user inputs, our
  approach optimizes the facial details against the trained
  classifiers, so as to synthesize a face which gives the desired
  personality impression. We demonstrate our approach for synthesizing
  3D faces giving desired personality impressions on a variety of 3D
  face models. Perceptual studies show that the perceived personality
  impressions of the synthesized faces agree with the target
  personality impressions specified for synthesizing the faces. Please refer to the supplementary materials for all results.

\end{abstract}

\vspace{-5mm}
\section{Introduction}
A face conveys a lot of information about a person. People usually
form an impression about another person in less than a second, mainly
by looking at another person's face. Researchers in psychology,
cognitive science, and biometrics conducted a lot of studies to
explore how facial appearances may influence personality
impression~\cite{willis2006first,hassin2000facing}. Some researchers investigated the relationship
between personality impressions and specific facial
features~\cite{eisenthal2006facial}. There are also
attempts in training machine learning models for predicting
personality impressions based on facial
features~\cite{gray2010predicting,joo2015automated}.

To create realistic 3D faces for the computer games, digital entertainments,  and
virtual reality applications, some works have been carried on generating realistic face~\cite{hu2017avatar}, vivid animation~\cite{Nicholas2018PVL}, natural expressions~\cite{marsella2013virtual}, and so on. Yet, synthesizing 3D faces that give certain personality is not explored, which is one of the most important considerations during the creative process. For example, the main characters in games and animations
are usually designed to look confident and smart, whereas the ``bad
guys'' are usually designed to look hostile. However, while there are
automatic tools for synthesizing human faces of different ethnicities
and genders, the problem of synthesizing 3D faces with respect to
personality impressions is still unsolved. We propose a data-driven
optimization approach to solve this problem.

%Creating 3D faces that give certain personality
%impressions is one of the most important considerations during the creative process of the computer games, animations,  and
%virtual reality applications for producing realistic virtual
%characters. For example, the main characters in games and animations
%are usually designed to look confident and smart, whereas the ``bad
%guys'' are usually designed to look hostile. However, while there are
%automatic tools for synthesizing human faces of different ethnicities
%and genders, the problem of synthesizing 3D faces with respect to
%personality impressions is still unsolved. We propose a data-driven
%optimization approach to solve this problem.

As the personality impression of a face depends a lot on its subtle
details, under the current practice, creating a face to give a
certain personality impression is usually done through a
``trial-and-error'' approach: a designer creates several faces; asks
for people's feedback on their impressions of the faces; and then
modifies the faces accordingly. This process iterates until a
satisfactory face is created. This design process involves substantial
tuning efforts by a designer and is not scalable. Manual creation of
faces could also be very challenging if the objectives are abstract or
sophisticated. For example, while it could be relatively easy to
create a face to give an impression of being friendly, it could be
hard to create a face to give an impression of being friendly but
silly, which could be desirable for a certain virtual character.

%\begin{figure*}
 %\centering
 % \includegraphics[width=0.85\linewidth]{figure-new/motivation2b}
  %\caption{What kind of roles should these people assume in a film?
   % Popular votes from a perceptual study: (a) friendly, (b) smart,
  %  (c) hostile, (d) unconfident, (e) silly, and (f) confident.}
%\label{fig:motivate} 
%\vspace{-6mm}
%\end{figure*}

We propose a novel approach to automate this face creation
process. Our approach leverages Convolutional Neural Networks (CNN)
techniques to learn the non-trivial mapping between low-level subtle
details of a face and high-level personality impressions. The trained
networks can then be applied for synthesizing a 3D face to give a
desired personality impression via an optimization process. We
demonstrate that our approach can automatically synthesize a variety
of 3D faces to give different personality impressions, hence
overcoming the current scalability bottleneck. The synthesized faces
could find practical uses in virtual world applications (\eg,
synthesizing a gang of hostile-looking guys to be used as enemies in a
game).

The major contributions of our paper include:
\begin{itemize}
\item Introducing a novel
problem of synthesizing 3D faces based on personality impressions.
\vspace{-1mm}
\item Proposing a learning-based optimization approach and  a data-driven MCMC sampler  for synthesizing
faces with desired personality impressions.

\item Demonstrating the practical uses of our approach for different novel face editing, virtual reality applications and digital entertainments.  

\end{itemize}
\vspace{-3mm}

\section{Related Work}
%Facial modeling is an important topic in computer graphics that has
%attracted much research attention. Common approaches fall into the two
%major categories of geometric
%manipulations~\cite{blanz1999morphable,ichim2015dynamic}
%and image
%manipulations~\cite{tian2016facial,le2011shape}. Please
%refer to a recent survey~\cite{salam2018survey} for a comprehensive
%review. We review some representative work closely relevant to our
%problem.
\textbf{Faces and personality impressions.}
%\label{sec:faces_and_personalities}
Personality impression is an active research topic in
psychology and cognitive science. Researchers are interested in studying how different
factors, \eg, face, body, profile, motion, influence the formation of
personality impression on
others~\cite{naumann2009personality}. Recent work~\cite{over2018spontaneous} suggests that facial appearances play an important
role in giving personality impressions.

%gosling2007personality,
  
Some works focused on examining what facial features influence
personality impression. 
%Dotsch and
%Todorov~\shortcite{dotsch2012reverse} found that significant
%information determining social impression is located near the mouth,
%eye, eyebrow, and hair regions.
% by using reverse correlation analysis
%techniques. 
%Pallett~\shortcite{pallett2010new} calculated a golden
%proportion of facial features based on a human face dataset, which
%consisted of face images that were evaluated as beautiful by
%human. 
Vernon et al.~\shortcite{vernon2014modeling} modeled the
relationship between physical facial features extracted from images
and impression of social traits. Zell et
al.~\shortcite{zell2015stylize} studied the roles of face geometry and
texture in affecting the perception of computer-generated faces. Some findings were adopted to predict human-related attributes based
on a face. Xu et al.~\shortcite{xu2015new} proposed a cascaded
fine-tuning deep learning model to predict facial
attractiveness. 
%Elorza et al.~\shortcite{elorza2017face} analyzed face
%beauty via manifold-based semi-supervised learning. Xie et
%al.~\shortcite{xie2015scut} proposed a benchmark dataset for analyzing
%facial beauty impression. 
Joo et al.~\shortcite{joo2015automated}
proposed an approach to infer the personality traits of a person from
his face.
%, such as ``smartness'', ``trustworthiness'' and
%``competence''. They showed how such traits could be used to
%predict the outcomes of real-world social events.

Motivated by these findings, we use deep learning techniques to learn
the relationship between facial appearances and personality
impressions based on a collected face dataset with personality
impression annotations, which is applied to guide the synthesis of 3D
faces to give desired personality impressions by an optimization.

\noindent\textbf{Face Modeling and Exaggeration.}
%\label{sec:faces_modeling}
Some commercial 3D modeling software can be used by designers for
creating 3D virtual characters with rich facial details, such as
Character Generator, MakeHuman, Fuse, and so on. These
tools provide a variety of controls of a 3D face model, including
geometry and texture, \eg, adjusting the shape of the nose, changing
skin color. However, to create or modify a face to give a certain
personality impression, a designer has to manually tune many low-level
facial features, which could be very tedious and difficult.

Another line of work closely relevant to ours is face exaggeration,
which refers to generating a facial caricature with exaggerated face
features.
%~\cite{brennan2007caricature,Suwajanakorn_2015_ICCV,tseng2012colored,le2011shape}.
Suwajanakorn et al.~\shortcite{Suwajanakorn_2015_ICCV} proposed an
approach for creating a controllable 3D face model of a person from a
large photo collection of that person captured in different
occasions. 
%Tseng et al.~\shortcite{tseng2012colored} proposed a
%statistics-based exaggerative (SBE) module to emphasize a person's
%most distinctive features on an input face image. The user can
%conveniently control the exaggeration rate.
 Le et al.~\shortcite{le2011shape} performed exaggeration differently by
using primitive shapes to locate the face components, followed by
deforming these shapes to generate an exaggerated face. They
empirically found that specific combinations of primitive shapes tend
to establish certain personality stereotypes. Recently, Tian and
Xiao~\shortcite{tian2016facial} proposed an approach for face
exaggeration on 2D face images based on a number of shape and texture
features related to personality traits.

Compared to these works, our learning-based optimization approach
provides high-level controls for 3D face modeling, by which designers
can synthesize faces with respect to specified personality impressions
conveniently.

\begin{figure}[t]
 \centering
 \vspace{-3mm}
  \includegraphics[width=1\linewidth]{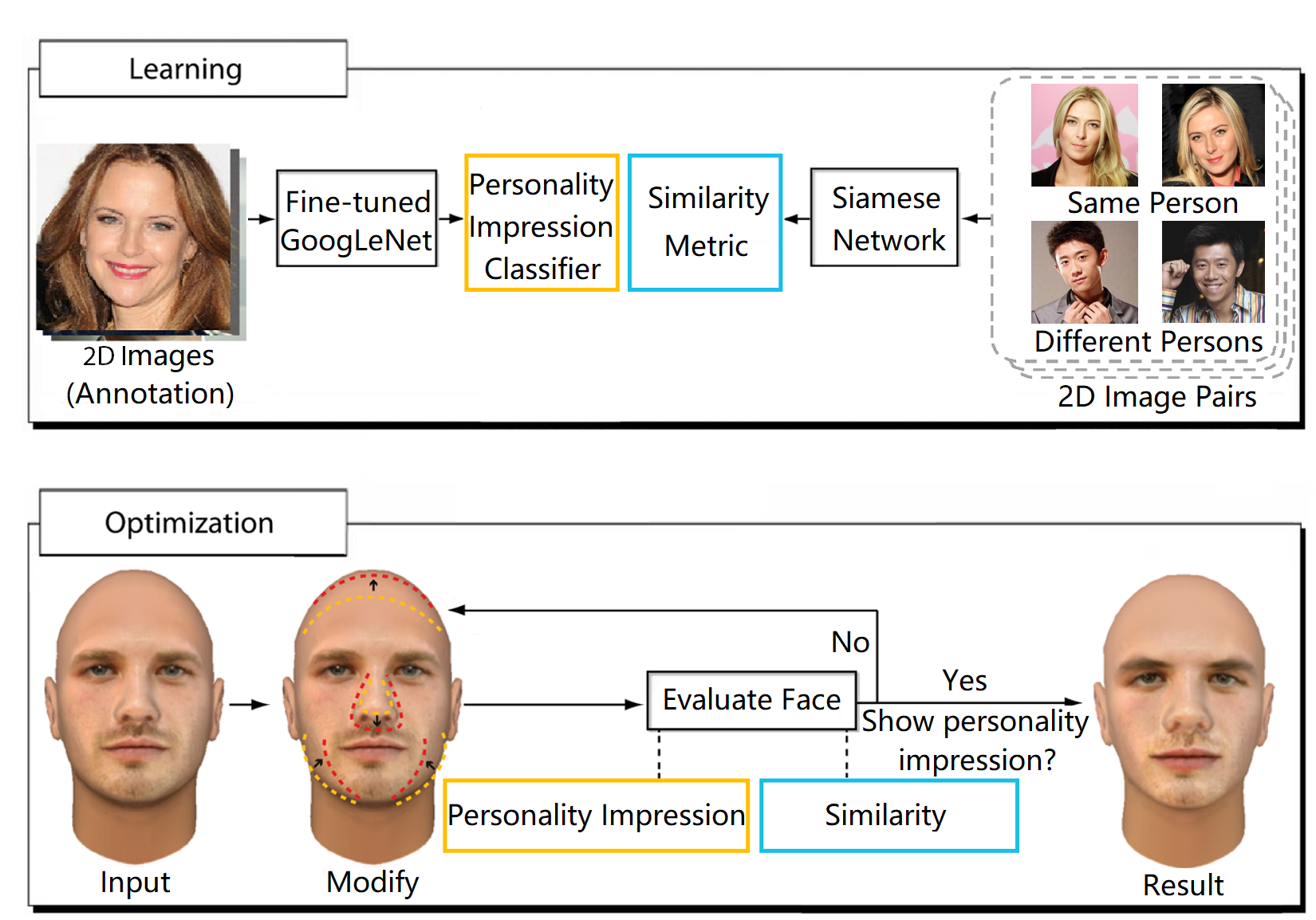}
  \parbox[t]{1.0\textwidth}{\relax}
\vspace{-5mm}
  \caption{\label{fig:framework}
           Overview of our approach.}
\vspace{-5mm}
\end{figure}

\noindent\textbf{Data-Driven 3D Modeling.}
%\label{sec:data_driven}
Data-driven techniques have been successfully applied for 3D modeling~\cite{kalogerakis2012probabilistic,talton2011metropolis}. 
%Lun
%et al.~\shortcite{lun2015elements} learned perceptual shape style
%similarity among 3D models. The trained classifiers were applied for
%style-based shape categorization and 3D model suggestion. 
Huang et al.~\shortcite{huang2017shape} devised
deeply-learned generative models for 3D shape synthesis. 
Ritchie et
al.~\shortcite{ritchie2015controlling} used Sequential Monte Carlo to
guide the procedural generation of 3D models in an efficient manner.
Along the direction of face modeling,
Saito~\shortcite{saito2016photorealistic} et al. used deep neural
networks trained with a high-resolution face database to automatically
infer a high-fidelity texture map of an input face image. 
%Han et
%al.~\shortcite{HanGY17} developed an intuitive sketch-based interface
%for creating 3D faces caricatures, where the inference of sketches is
%performed by a deep regression network.

%Face appearances carry a lot of subtle details. To represent a 3D face
%with details, the parameter space is generally
%high-dimensional. 
Modeling the relationships between low-level facial
features and high-level personality impressions is difficult. In
addition, directly searching in such a complex and high-dimensional
space is inefficient and unstable. In our work, we apply data-driven
techniques to model the relationship between facial appearances and
personality impressions. Furthermore, we speed up face synthesis by
formulating a data-driven sampling approach to facilitate the
optimization.
 
\vspace{-2mm}
\section{Overview}
\label{sec:overview}

Figure~\ref{fig:framework} shows an overview of our approach. Given an
input 3D face, our approach optimizes the face geometry and texture
such that the optimized face gives the desired personality impression
specified by the user. To achieve this goal, we present an automatic
face synthesis framework driven by personality impression, which consists of two
stages: {\bf learning} and {\bf optimization}.

In the learning stage, we define $8$ types of personality
impression. Then we learn a CNN personality impression classifier for
each type. 
To train the CNN classifier, we collected $10,000$
images from CASIA WebFace database ~\cite{yi2014learning} and
annotated them with the corresponding personality impression.  
We also
learn an end-to-end metric to evaluate the similarity between the
synthesized face and the input one. The metric plays the role of
constraining 3D face deformation. 

In the optimization stage, our approach modifies the face geometry and
texture iteratively. The resulting face is then evaluated by the
personality impression cost function, as well as the similarity cost
function. To speed up our optimization,
%approach such that it can be applied for interactive face modeling, 
we devise a data-driven sampling approach based on the learned
priors. The optimization continues until a face giving the desired
personality impression is synthesized.
%The classifier and the similarity
%metric construct a cost function subsequently, where the classifier
%guides an optimizer to synthesize a 3D face giving the desired
%personality impression, and the similarity metric constrains the
%synthesized face to keep the appearance consistency with the input
%one.  A trade-off parameter is utilized to balance the influence of
%the personality impression and the similarity during the optimization.

%To search the solution space efficiently, we learn a prior
%distribution of face appearance for each personality impression from
%the annotated dataset, and a prior distribution of manual manipulation
%on different face regions from artists. 

\vspace{-2mm}

\section{Problem Formulation}
\textbf{Personality Impression Types.}  
In our experiments, we use four pairs of personality impressions types:
a) smart/silly; b) friendly/hostile; c) humorous/boring; and d)
confident/unconfident. These personality impression types are commonly used in psychology~\cite{mischel2013personality}.

\noindent\textbf{3D Face Representation.}
\label{sec:3drep}
To model a 3D face, we use a multi-linear PCA approach to represent
the face geometry and the face texture based on the morphable face
models~\cite{blanz1999morphable}, akin to the representation
of~\cite{hu2017avatar}. Our approach operates on a textured 3D face mesh model. We represent a face $(\mathbf{V}, \mathbf{T})$ by its geometry
$\mathbf{V} \in \mathbf{R}^{3n}$, which is a vector containing the 3D
coordinates of the $n=6,292$ vertices of the face mesh, as well as a
vector $\mathbf{T} \in \mathbf{R}^{3n}$ containing the RGB values of
the $n$ pixels of its texture image.

%\setlength{\columnsep}{12pt}
%\begin{wrapfigure}[12]{r}{0.3\linewidth}
%\vspace{-3mm}
%  \centering
%{\includegraphics[width=0.9\linewidth]{figure-new/facialregion-45}}\hfill
%\vspace{-3mm}
%\caption{\label{fig:facialregion} Face regions. \vspace{-5mm}}
%\end{wrapfigure}

Each face is divided into $8$ regions (eyes, jaw, nose, chin, cheeks,
mouth, eyebrows and face contour) as depicted in the supplementary materials. For each face region, we learn two Principal
Component Analysis (PCA) models for representing its geometry and
texture in low-dimensional spaces. The PCA models are learned using 3D
faces from the Basel Face Model database~\cite{bfm09}.

 First, we
manually segment each face into the eight regions. Then, for each
region, we perform a PCA on the geometry and a PCA on the texture to
compute the averages and the sets of eigenvectors. In our
implementation, when doing the PCAs for the $r$-th region, for all
vertices in $\mathbf{V}$ and all pixels in $\mathbf{T}$ that do not
belong to the $r$-th region, we just set their values to zero so that
all regions have the same dimensionality and can be linearly combined
to form the whole face $(\mathbf{V}, \mathbf{T})$: \vspace{-2mm}
%\begin{align}
%\label{eqn:facemodel}
%\mathbf{V}&=\sum_{r=1}^8(\bar{\mathbf{V}}_r+  {\Lambda}_r {\mathbf{v}_r}),\\
%\mathbf{T}&=\sum_{r=1}^8(\bar{\mathbf{T}}_r+  {\Gamma}_r {\mathbf{t}_r}).
%\end{align}
{\small
\begin{equation}
\label{eqn:facemodel}
\mathbf{V}=\sum_{r=1}^8(\bar{\mathbf{V}}_r+  {\Lambda}_r {\mathbf{v}_r}),\,
\mathbf{T}=\sum_{r=1}^8(\bar{\mathbf{T}}_r+  {\Gamma}_r {\mathbf{t}_r}).
\end{equation}
} \vspace{-3mm}

Here $r$ is the index of a face region;
$\bar{\mathbf{V}}_r \in \mathbf{R}^{3n}$ and
$\bar{\mathbf{T}}_r \in \mathbf{R}^{3n}$ denote the average geometry
and average texture for the $r$-th face region;
${\Lambda}_r \in \mathbf{R}^{3n \times m}$ and
${\Gamma}_r \in \mathbf{R}^{3n \times m}$ are matrices whose columns
are respectively the eigenvectors of the geometry and texture. We use
$m=40$ eigenvectors in our experiments.
$\mathbf{v}_r \in \mathbf{R}^{m}$ and
$\mathbf{t}_r \in \mathbf{R}^{m}$ are vectors whose entries are the
coefficients corresponding respectively to the eigenvectors of the
geometry and texture. This representation allows our approach to
manipulate an individual face region by modifying its coefficients
$\mathbf{v}_r$ and $\mathbf{t}_r$. Based on the PCA models of the $8$ face regions, a 3D face $(\mathbf{V}, \mathbf{T})$ is parameterized as a tuple
$\theta=(\textbf{v}_1,\textbf{v}_2,\cdots,\textbf{v}_8,\textbf{t}_1,\textbf{t}_2,\cdots,\textbf{t}_8)$ containing the coefficients.

\noindent\textbf{Facial Attributes.}
\label{sec:fac}
Although different faces can be synthesized by changing the face
coefficients $\textbf{v}_i$ and $\textbf{t}_i$, in general these
coefficients do not correspond to geometry and texture facial
attributes that can be intuitively controlled by a human modeler for
changing a face's outlook. It would be desirable to devise a number of
facial attributes in accordance with human language (\eg, ``changing
the mouth to be wider''), to facilitate designers in interactively
modifying a 3D face, and to allow our optimizer to learn from and
mimic human artists on the tasks of modifying a face with respect to
personality impression.

We describe how the effect of changing a facial attribute $a$ can be
captured and subsequently applied for modifying a face. For
simplicity, we assume that each facial attribute is defined only in
one face region rather than across regions. Based on a set of exemplar
faces $\{(\mathbf{V}_i, \mathbf{T}_i)\}$ from the Basel Face Model
database with assigned facial attribute $a$, we compute the sums:
%{\small
%\begin{align}
%\Delta \mathbf{V}_a &= \frac{1}{A} \sum_{i=1} \mu_i (\mathbf{V}_i-\bar{\mathbf{V}}),
%\Delta \mathbf{T}_a &= \frac{1}{A} \sum_{i=1} \mu_i (\mathbf{T}_i-\bar{\mathbf{T}}),
%\label{eqn:rep}
%\end{align}
%}
\vspace{-2mm}
{\small
\begin{equation}
\Delta \mathbf{V}_a = \frac{1}{A} \sum_{i=1} \mu_i (\mathbf{V}_i-\bar{\mathbf{V}}),\,
\Delta \mathbf{T}_a = \frac{1}{A} \sum_{i=1} \mu_i (\mathbf{T}_i-\bar{\mathbf{T}}),
\label{eqn:rep}
\end{equation}
}\vspace{-3mm}

where $\bar{\mathbf{V}}$ and $\bar{\mathbf{T}}$ are the average
geometry and average texture computed over the whole Basel Face Model
dataset. $\mu_i \in [0,1]$ is the markedness of the attribute in face
$(\mathbf{V}_i,\mathbf{T}_i)$, which is manually assigned. $A=\sum_{i=1}\mu_i$ is
the normalization factor. Given a face $(\mathbf{V}, \mathbf{T})$, the result of changing facial
attribute $a$ on this face is given by
$(\mathbf{V} + \beta \Delta \mathbf{V}_a, \mathbf{T} + \beta \Delta
\mathbf{T}_a)$,
where $\beta$ is a parameter for controlling the extent of applying
facial attribute $a$.

In total, we devise $160$ facial attributes. Each attribute is modeled
by $5$ example faces. We demonstrate the effect of each
attribute on an example face. It is worth noting that the representation of a 3D face can be replaced by other 3D face representations that provide controls of a face. 
%To
%demonstrate how our approach can be generally applied for other face
%modellers, we incorporated our approach into different face
%representation and modeling tools (\eg, 3DMM, MakeHuman, Character
%Creator, and Adobe Fuse). 
Please find the corresponding results in the supplementary material.

\noindent\textbf{Optimization Objectives.}
\label{sec:objective}
We synthesize a 3D face to give a desired personality impression by an
optimization process, which considers two factors: (1) \textbf{Personality Impression}: how likely the synthesized face gives the desired personality impression.
(2) \textbf{Similarity Metric}: how similar the synthesized face is with the input face.

% {\bf Personality Impression} (how likely the synthesized face shows the desired personality impression) and {\bf Similarity Metric} (how similar the synthesized face is with the input face).

%\begin{itemize}
%\item {Personality Impression}: how likely the synthesized face gives the desired personality impression.
%\item {Similarity Metric}: how similar the synthesized face is with the input face.
%\end{itemize}

Given an input 3D face and a desired personality impression type, our
approach synthesizes a 3D face which gives the desired personality
impression by minimizing a total cost function:
\begin{equation}
\label{eqn:cost}
\mathbf{C}(\theta)= \mathbf{C_{\textrm{p}}}(I_{\theta},P)+\lambda \mathbf{C_{\textrm{s}}}(I_\theta,I_{\textrm{i}}),
\end{equation}
where
$\theta=(\textbf{v}_1,\textbf{v}_2,\cdots,\textbf{v}_8,\textbf{t}_1,\textbf{t}_2,\cdots,\textbf{t}_8)$
contains the face coefficients for synthesizing a 3D
face. $\mathbf{C_{\textrm{p}}(\cdot)}$ is the personality impression
cost term for evaluating image $I_{\theta}$ of the face synthesized
from $\mathbf{\theta}$ with regard to the desired personality
impression type $P$. The face image is rendered using the frontal view
of the face. Lambertian surface reflectance is assumed and the
illumination is approximated by second-order spherical
harmonics~\cite{ramamoorthi2001efficient}.
$\mathbf{C_{\textrm{s}}(\cdot)}$ is the similarity cost term, which
measures the similarity between the image $I_\theta$ of the
synthesized face and the image $I_{\textrm{i}}$ of the input face,
constraining the deformation of the input face during the
optimization. $\lambda$ is a trade-off parameter to balance the costs
of personality impression and similarity.

\vspace{-2mm}
 
 \section{Personlaity Impression Classification}

To compute the personality impression cost $\mathbf{C_{\textrm{p}}}$
for a synthesized face in each iteration of the optimization, we
leverage modern deep CNN with high-end performances and train a
classifier for each personality impression type, which provides a
score for the synthesized face with regard to the personality
impression type. To achieve this, we create a face image
dataset annotated with personality impression labels based on CASIA WebFace
 database~\cite{yi2014learning}, which consists of $10,000$ face
 images covering both genders and different ethnicities. Then, we
fine-tune GoogLeNet~\cite{szegedy2015going} with a personality
impression classification task on the dataset. Please refer to our supplementary material for more details about the database.

\textbf{Learning.}\label{sec:learning}
We construct our network based on the original GoogLeNet with
pre-trained parameters. The network is $22$ layers deep with $5$
average pooling layers. It has a fully connected layer with $1,024$
units and rectified linear activation. During the fine-tuning process, the images with the
corresponding labels in the personality impression dataset are fed to
the network and an average classification loss is applied. Please find the details about the training process and the visualization of the network in our supplementary material.

%We use a GPU-based engine and implement asynchronous stochastic
%gradient descent with $0.9$ momentum in the fine-tuning stage and a
%fixed learning rate schedule (with learning rate decreased by $4\%$
%every $8$ epochs). The mini-batch size is $128$. Thus, the original
%GoogLeNet model is fine-tuned to adapt to a personality impression
%classification task. 
%Figure~\ref{fig:cnn} shows some classification
%examples.  

%After fine-tuning, each face image of our dataset is propagated to the
%fine-tuned network, and a $1,024$-dimension feature vector
%$\mathbf{g}$ is extracted from the $22$-nd layer. Based on those
%feature vectors, our approach learns a linear SVM classifier for each
%personality impression type. 

\textbf{Evaluation.}
We evaluate our approach with real face images on personality
impression classification. We compare the fine-tuned GoogLeNet of our
approach ({\bf CNN-R}) with the approach of using landmark
features~\cite{zhu2012face} and being trained by a standard SVM
classifier ({\bf LM-R}). Both approaches are based on the same splitting 
strategy of the dataset ($70\%$ for training and $30\%$ for testing). CNN-R 
attains an average accuracy of $86.9\%$ across all personality impression types,
whereas LM-R attains an average accuracy of $75.3\%$.
Please refer to the supplementary material for more quantitative comparison
results.

\vspace{-2mm}
\section{Face Similarity Metric}
\label{sec:similar}

To constrain the synthesized face to look similar to the
input face, we evaluate the similarity between the image $I_\theta$ of
the synthesized face and image $I_{\textrm{i}}$ of the original input
face in the optimization. To achieve this, we train
a Siamese network~\cite{chopra2005learning}, an end-to-end network, to
evaluate whether a pair of face images correspond to the same
face. The network learns a feature extractor which takes face images
and outputs feature vectors, such that the feature vectors
corresponding to images of the same face are close to each other,
while those corresponding to images of different faces are far away
from each other.

%\newcommand{\tabincell}[2]{\begin{tabular}{@{}#1@{}}#2\end{tabular}}
 %\begin{table}[t]
 %\centering
 % \vspace{1mm}
 	
% \caption{ Accuracy of personality impression classification with fine-tuned GoogLeNet %features ({\bf CNN}) and landmark features ({\bf LM}) over real data ({\bf R}) and %synthesis data (\bf{S}).}
   %\begin{tabular}{ l c c c }
 %	\cline{1-4}
 %	\textbf{Category} & \textbf{CNN-R} & \textbf{LM-R}& \textbf{CNN-S} \\
 %	\hline
 %	Friendly & 91\%& 82\%  & 94\%  \\
 %	Hostile  & 87\%& 75\%  & 92\% \\
 %	Smart & 87\% & 73\%  & 88\%  \\
 %	Silly  & 86\% & 80\%  & 92\%\\
 %	Humorous & 85\%& 81\%  & 89\%  \\
 %	Boring & 89\%& 70\%  & 87\%  \\
 %	Confident & 86\% & 74\% & 83\%  \\
 %	Unconfident & 84\% & 68\% & 79\% \\
  % \hline
  % \end{tabular} 
 %  \vspace{-2mm}
 	
 %	\label{tbl:evaluation}
 %	\vspace{-4mm}
 %\end{table}

% To evaluate the similarity between two faces, we use a Siamese
% network~\cite{chopra2005learning}, an end-to-end network, to evaluate the
% similarity metric between the images of the synthesized face $I_\theta$
% and the original input face $I_{\textrm{i}}$ for each iteration of the
% optimization. The goal of the network is to learn a feature extractor,
% which produces output feature vectors \craig{not sure what this means}
% that are nearby for the images from the same category, and far away
% for the ones from different categories.

We train the Siamese network using the LFW
dataset~\cite{huang2007labeled}. The training dataset is constructed
as $\{(I_{\textrm{a}},I_{\textrm{b}},l)\}$, where $I_{\textrm{a}}$ and
$I_{\textrm{b}}$ are any two images from the LFW dataset, and $l$ is
the label. If $I_{\textrm{a}}$ and $I_{\textrm{b}}$ are from the same
face, $l=1$, otherwise $l=0$. 

The Siamese network consists of two identical Convolutional Networks
that share the same set of weights $W$. The training process learns
the weights $W$ by minimizing a loss function $\mathcal{L}=l\mathcal{L}_1+(1-l)\mathcal{L}_2$, where $\mathcal{L}_1=\lVert G_W(I_{\textrm{a}})-G_W(I_{\textrm{b}})\lVert$ and
$\mathcal{L}_2= \max(0, \rho-\lVert G_W(I_{\textrm{a}})-G_W(I_{\textrm{b}}) \lVert)$.
%\begin{align}
%\mathcal{L}&=l\mathcal{L}_1+(1-l)\mathcal{L}_2, \nonumber \\
%\mathcal{L}_1&=\lVert G_W(I_{\textrm{a}})-G_W(I_{\textrm{b}})\lVert, \nonumber\\
%\mathcal{L}_2&= \max(0, \rho-\lVert G_W(I_{\textrm{a}})-G_W(I_{\textrm{b}}) \lVert),\nonumber
%\end{align}
$G_W(I)$ is the mapped features of an input face image $I$,
which are synthesized by the learned identical Convolutional
Network. By minimizing the loss function $\mathcal{L}$, the distance
between the mapped features of $I_{\textrm{a}}$ and $I_{\textrm{b}}$
is driven by $\mathcal{L}_1$ to be small if $I_{\textrm{a}}$ and
$I_{\textrm{b}}$ correspond to the same face, and is driven by
$\mathcal{L}_2$ to be large vice versa. The constant $\rho$ is set as
$2.0$. The parameters are learned by standard cross entropy loss and
back-propagation of the error.

\vspace{-2mm}
\section{Cost Functions}
Given a textured 3D face model and a desired personality impression
type as the input, our approach employs a data-driven MCMC sampler to
update the face coefficients $\theta$ iteratively so as to
modify the face. In each iteration, the synthesized face represented
by $\theta$ is evaluated by the total cost
$\mathbf{C}(\cdot)=
\mathbf{C_{\textrm{p}}}(\cdot)+\lambda\mathbf{C_{\textrm{s}}}(\cdot)$.
The optimization continues until a face giving the desired
personality impression is synthesized. We discuss the personality
impression cost $\mathbf{C_{\textrm{p}}}$ and the similarity cost
$\mathbf{C_{\textrm{s}}}$ in the following.

\noindent\textbf{Personality Impression Cost.}
The image $I_{\theta}$ of the face synthesized by face coefficients
$\theta$ is evaluated with respect to the desired personality
impression type $P$ in the cost function $\mathbf{C_{\textrm{p}}}$,
defined based on the fine-tuned GoogLeNet:
\begin{align}
\label{eqn:pcost}
\mathbf{C_{\textrm{p}}}(I_{\theta},P)=1-\frac{\textrm{exp}(x_1)}{\textrm{exp}(x_1)+\textrm{exp}(x_2)},
\end{align}
where $[x_1,x_2]^T=\mathbf{w}_P^T \mathbf{g}$ is the output of the
full connected layer of the fine-tuned network. $x_1$ and $x_2$
reflect the possibilities of the image $I_\theta$ belonging to the
personality impression type $P$ or not, respectively.
$\mathbf{g} \in \mathbf{R}^{1,024}$ is the face feature vector of
$I_{\theta}$ on the $22$-nd layer of the network;
$\mathbf{w}_P \in \mathbf{R}^{1,024 \times 2}$ contains the parameters
of the full connected layer, which map the feature vector $\mathbf{g}$
to a 2D vector (our fine-tuned network is a two-category classifier).

A low cost value means the synthesized face image gives the desired
type of personality impression, according to the classifier trained by
face images annotated with personality impression labels.

%\begin{equation}
%\label{eqn:pcost}
%\mathbf{C_{\textrm{p}}}(P,I_{\theta})=-{\mathbf{w}_P}^T \mathbf{g}+b, 
%\end{equation}
%
%where $\mathbf{w}_P$ and $b$ are the parameters learned by the standard linear SVM approach; $\mathbf{g}$ is the face feature vector of $I_{\theta}$ extracted from the fine-tuned GoogLeNet. A low cost value means the synthesized face image shows the desired type of personality impression, according to the classifier trained by face images annotated with personality impression labels. \craig{is the cost within 0 to 1? if not, should normalize}

\begin{figure}[t]
\vspace{-3mm}
  \centering
\subfloat[Input]{\includegraphics[width=0.25\linewidth]{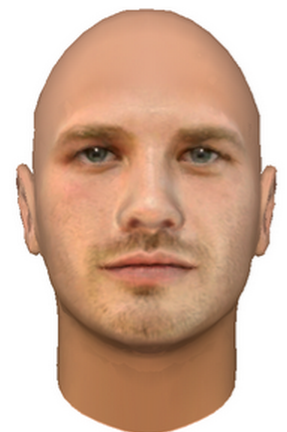}}\hspace{2mm}
\subfloat[$\lambda=0.3$]{\includegraphics[width=0.25\linewidth]{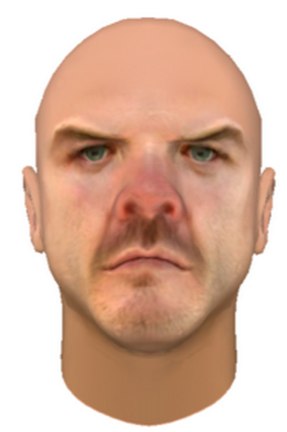}}\hspace{2mm}
\subfloat[$\lambda=0.5$]{\includegraphics[width=0.25\linewidth]{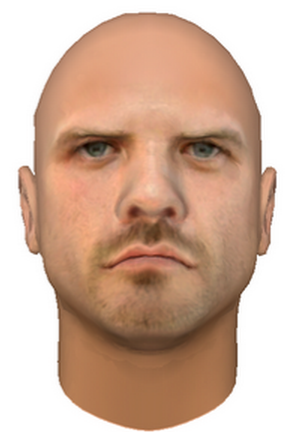}}\\
\vspace{-3mm}
\caption{Effects of $\lambda$ when opimizing an example face to give a hostile personality impression. A larger $\lambda$ constrains the
  synthesized face to resemble the input face more closely.}
  \vspace{-4mm}
  \label{fig:similarity}
\end{figure}

\noindent\textbf{Similarity Cost.}
%$\mathbf{\Psi(F,s)}$ is the similarity cost term for evaluating the
%similarity between the synthesized face $\mathbf{s}$ and the input
%face $\mathbf{F}$, which penalizes deformations during
%optimization. 
We want to constrain the synthesized face to look similar to the input
face. To achieve this, we apply the Siamese network trained for evaluating the similarity
between a pair of face images to define a similarity cost as a soft
constraint of the optimization:
\begin{equation}
\label{eqn:simalrityd}
\mathbf{C_{\textrm{s}}}(I_{\theta},I_{\textrm{i}})= \frac{1}{G} \lVert G_W(I_{\theta})-G_W(I_{\textrm{i}})\lVert,
\end{equation}
where $G_W(I_{\theta})$ and $G_W(I_{\textrm{i}})$ are the feature
vectors of the image $I_{\theta}$ of the synthesized face and the
image $I_{\textrm{i}}$ of the input face computed by the Siamese
network. $G = \max(\{ \lVert G_W(I) - G_W(I_{\textrm{i}}) \lVert\})$
is a normalization factor computed over all face images $I$ from the
LFW dataset. A low cost value means that the synthesized face image
$I_{\theta}$ is similar to the input face image $I_{\textrm{i}}$.

% the two faces of $I_{\theta}$ and $I_{\textrm{i}}$ are similar.

To demonstrate how the similarity cost term affects the face synthesis
results during optimization, we show an example of optimizing a face
model with the personality impression type of hostile in
Figure~\ref{fig:similarity}. When the trade-off parameter $\lambda$ is
set as $0.3$, the face is optimized to become more hostile-looking yet
it differs from the input face significantly. When $\lambda$ is set as
$0.5$, the face is optimized to look somewhat hostile and it resembles
the input face more closely. In our experiments, we set $\lambda=0.5$
by default.

% We use the learned Siamese network to compute a feature vector for
% any input image. Then the similarity cost between two images are
% defined as the distance between their output features:

\vspace{-2mm}
\section{Face Synthesis by Optimization}
\label{sec:ddmcmc}

%\label{sec:ddmcmc}
We use a Markov chain Monte Carlo (MCMC) sampler to explore the space
of face coefficients efficiently. As the top-down nature of MCMC
sampling makes it slow due to the initial "burn-in" period,
%, some
%approaches~\cite{jampani2015informed,kulkarni2015picture}
%used a bottom-up strategy which improves the speed and accuracy of
%MCMC sampling to help convergence. Inspired by these bottom-up
%approaches, 
we devise a data-driven MCMC sampler for our problem. 
%We
%show in our experiments (see Section~\ref{sec:user_control} and video
%demos) that with the speedup attained by our data-driven MCMC sampler,
%we are able to apply our optimization to realize an interactive face
%modeling tool. 
We propose two types of data-driven Markov chain
dynamics: Region-Move and Prior-Move, corresponding to local
refinement and global reconfiguration of the face.
 
% \craig{Prior-Move involves a bigger jump and should be applied just
%   occasionally to jump out of local minimum}

\ssection{Region-Move.} We want to learn from how human artists modify
faces to give a certain personality impression, so as to enable our
sampler to mimic such modification process during an
optimization. Considering that each face region's contribution to a
specified personality impression is different, we devise a Region-Move
which modifies a face according to ``important'' face regions likely to be
associated with the specified personality impression in
training data.

Our training data is created based on $5$ face models. We recruited
$10$ artists who are familiar with face modeling (with $5$ to $10$
years of experience in avatar design and 3D modeling). 
Each artist was asked to modify each of the $5$ face
models to give the $8$ personality impression types by controlling the
facial attributes. After the
manual modifications, we project the original $5$ face models and all
the manually modified face models into the PCA spaces, so that each face can be represented by its
face coefficients $\theta$.

% We recruited ten artists who are familiar with face modeling (with $5$
% to $10$ years of experience in avartar design and 3D modeling). They
% are requested to adjust $10$ face models manually to show the $8$
% personality impression types. After the adjustment, we project the
% original 10 face models and all manually modified models into the PCA
% face space, and compute the change between them.

For each personality type, let
$\Delta \theta=(\Delta \textbf{v}_1,\cdots,\Delta \textbf{v}_8,\Delta
\textbf{t}_1,\cdots,\Delta \textbf{t}_8)$
contain the sums of face coefficients differences for the $8$ face
regions.
$\Delta \textbf{v}_r = \sum || \textbf{v}_r - \textbf{v}_r' ||$ is the
sum of differences of the geometry coefficients of the $r$-th face
region, where $\textbf{v}_r$ and $\textbf{v}_r'$ are the geometry
coefficients of the original face model and a face model modified by
an artist respectively. The sum of differences of the texture
coefficients $\Delta \textbf{t}_r$ is similarly defined. Suppose the current face is $(\mathbf{V}, \mathbf{T})$ with face
coefficients $\theta$. During sampling, a face region $r$ is selected
with probability
$0.5 \frac{\Delta \textbf{v}_r }{\sum \Delta \textbf{v}_i}+0.5
\frac{\Delta \textbf{t}_r}{\sum \Delta \textbf{t}_i}$.
Then a facial attribute $a$ in face region $r$ is randomly selected
and modified so as to create a new face
$(\mathbf{V} + \beta \Delta \mathbf{V}_a, \mathbf{T} + \beta \Delta
\mathbf{T}_a)$
with new face coefficients $\theta'$, where $\beta \sim U(-1.0,1.0)$.
The changes $\Delta \mathbf{V}_a$ and $\Delta \mathbf{T}_a$ are
learned in Section Facial Attribute for each facial attribute
$a$. 

Essentially, a face region that is more commonly modified by artists
to achieve the target personality impression type is modified by our
sampler with a higher probability. 
%Note that our sampler does not
%directly select a facial attribute with a high modification
%probability (as observed from training data) to modify; this strategy
%would have suffered from overfitting because the size of the training
%data is small given there are many ($160$) facial attributes.

% Note that our sampler selects a region having a high modification
% probability and then selects a face attribute within that region to
% modify, rather than directly selecting a face attribute having a
% high modification probability to modify. To avoid overfitting, the
% latter strategy would have required a lot of training data because
% there are many ($160$) facial attributes.

\begin{figure}[t]
 \centering
 \vspace{-1mm}
  \includegraphics[width=1\linewidth]{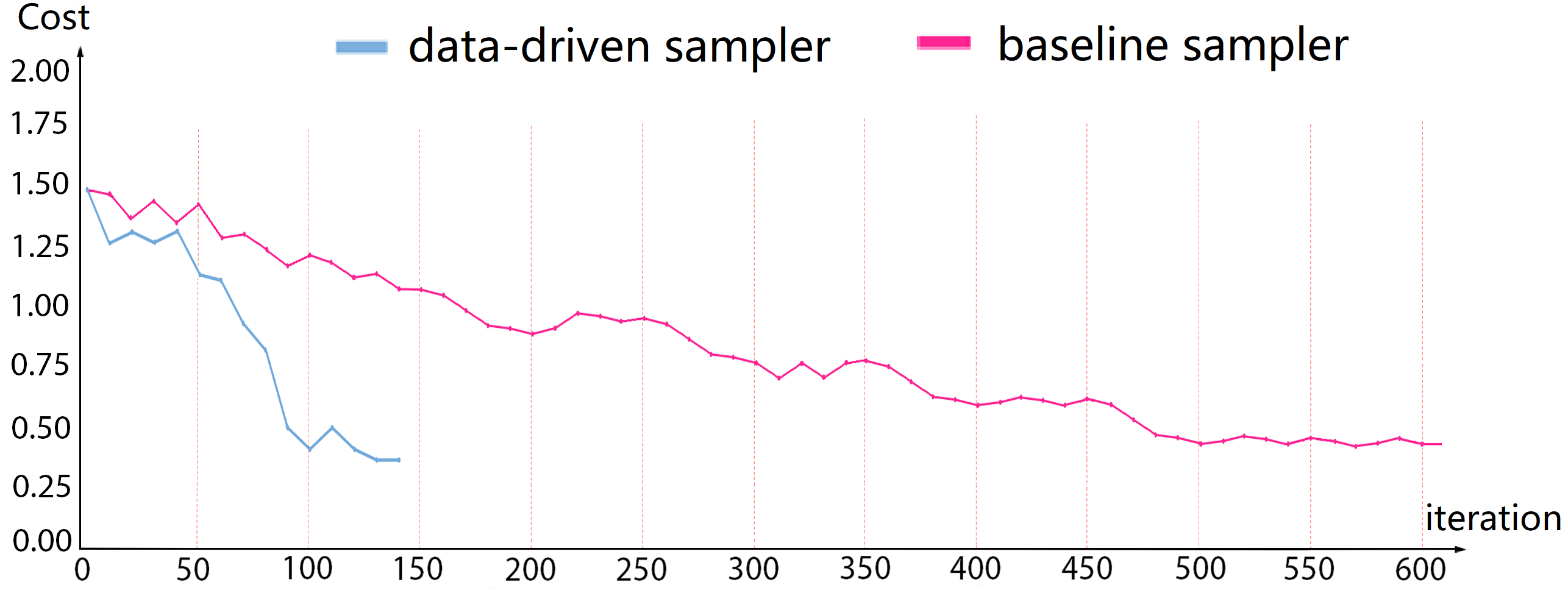}
  %\parbox[t]{0.3\columnwidth}{\relax}
      \vspace{-5mm}
  \caption{Total costs over iterations in optimizing a face using a
    data-driven sampler and a baseline sampler.}
    \vspace{-5mm}
\label{fig:optimization}
\end{figure}

% For each personality impression type, denote the parameter change of
% face geometry in each face region as
% $\Delta V=\{\Delta V_1, \Delta V_2,\cdots,\Delta V_8\}$. $\Delta V_i$
% is the parameter change of the $i$-th region and
% $\Delta V_i= \sum ||\Delta v_i||$. $\Delta v_i = ||v_i-v_i'||$ is the
% difference between the coefficients of the original model $v_i$ and of
% the modified model by an artist $v_i'$. Then a Gaussian distribution
% is used to model $\Delta V$.

% Similarly, we can get a Gaussian distribution on the parameter change
% of face texture $\Delta T$.

% During the optimization, when the sampler choose to perform
% Region-Move by the probability $\alpha$, the distribution guides the
% sampler to spend more time to explore important face regions. \craig{how exactly?}

\ssection{Prior-Move.}  We also leverage the personality impression
dataset to learn a prior distribution
of the face coefficients $\theta$ for each personality impression,
so as to guide our sampler to sample face coefficients near the prior
face coefficients, which likely induce a similar personality
impression.

\begin{figure*}[t]
 \centering
  \vspace{-3mm}
  \includegraphics[width=0.98\linewidth]{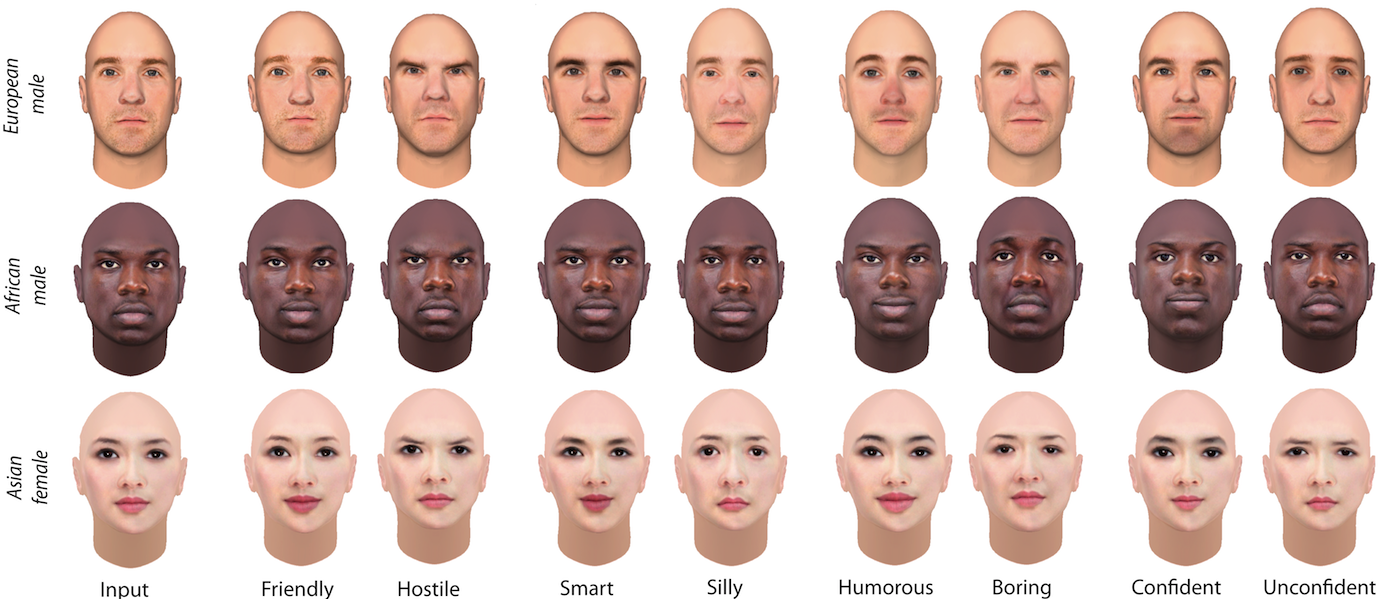}
\vspace{-3mm}
  \caption{Results of synthesizing faces with different personality impression types.}
\label{fig:mass}
 \vspace{-5mm}
\end{figure*}

For each personality impression type $P$, we estimate a prior
distribution with the following steps:

\begin{itemize}
\item[(1)] Select images in the personality impression dataset which
  are annotated with the personality impression type $P$; and form a
  subset $D_P=\{I_d\}$.

\item[(2)] Reconstruct the corresponding 3D face model for each image
  $I_d \in D_P$ by the implementation
  of~\cite{blanz1999morphable,blanz2003face}. These 3D face models are
  projected onto the PCA spaces and are represented using face coefficients. Thus, we form a face
  coefficients set $\Theta_P=\{\theta_d\}$.
\item[(3)] Fit a normal distribution for each of the geometry and
  texture coefficients ($\mathbf{v}_r$ and $\mathbf{t}_r$) of each
  face region $r$ based on $\Theta_P$.
\end{itemize}

Given the prior distribution, our sampler draws a value from the
normal distribution of each of the geometry and texture coefficients,
to generate new face coefficients $\theta'$. 

% During sampling, based on the current face coefficients $\theta$, the
% sampler draws new face coefficients $\theta'$ from the prior
% distribution. First, it finds $\hat{\theta} \in \Theta_P$ nearest to
% $\theta$. Then, it draws new face coefficients
% $\theta' \sim \mathcal{N}(\hat{\theta}, \hat{\sigma}^2)$, where
% $\hat{\sigma}$ is the standard deviation of the kernal associated with
% $\hat{\theta}$. \craig{theta is a tuple, not a vector, can do kernel estimation on tuple? mathematically ok?}

\noindent\textbf{Optimization.} 
We apply simulated annealing with a Metropolis-Hastings state-searching
step to search for face coefficients $\theta$ that minimize the total cost function $C$.
In each iteration of the optimization, one type of moves is selected
and applied to propose new face coefficients $\theta'$, which is
evaluated by the total cost function $\mathbf{C}$. The Region-Move and
Prior-Move are selected with probabilities $\alpha$ and $1-\alpha$
respectively. In our experiments, we set $\alpha=0.8$ by default.
% to
%slightly favor selecting Region-Move which corresponds to more local
%refinement of the face. 
The proposed face coefficients $\theta'$
generated by the move are accepted according to the Metropolis
criterion:
\begin{align}
Pr(\theta'|\theta) &= \min \bigg\{ 1,\frac{f(\theta')}{f(\theta)}\bigg\},
\end{align}
where $f(\theta) = {\exp}^{-\frac{1}{t}C(\theta)}$ is a Boltzmann-like
objective function and $t$ is the temperature parameter of the
annealing process. By default, we empirically set $t$ to $1.0$ and
decrease it by $0.05$ every $10$ iterations until it reaches zero. We
terminate the optimization if the absolute change in the total cost
value is less than $5\%$ over the past $20$ iterations. In our
experiments, a full optimization takes about $100-150$ iterations
(about $15$ seconds) to finish.

%Here, $\mathbf{C}(\cdot) = \mathbf{C}_P(\cdot) + \lambda \mathbf{C}_S(\cdot)$.

Figure~\ref{fig:optimization} shows an example of optimizing a face
using the proposed data-driven sampler and using a baseline sampler
which randomly picks one of the facial attributes and resets its value
randomly. By using more effective moves, the data-driven optimization
converges faster to obtain a solution with a lower cost value.

\vspace{-5pt}
\section{Experiments}
\label{sec:experiments}
We conducted experiments on a Linux machine equipped with an Intel
i7-5930K CPU, 32GB of RAM and a Nvidia GTX 1080 graphics
card. The optimization and learning components of our approach were
implemented in C++. 

%In our paper, we created a number of datasets to drive our approach, including: personality impression dataset (Section~\ref{sec:face_dataset}), face similarity dataset (Section~\ref{sec:similar}), etc.. All datasets are summarized in the \emph{supplementary material}s. We will release our datasets publicly.

%\begin{figure*}[t]
%  \centering
%{\includegraphics[width=0.5\linewidth]{figure/new/mixture_male1}}\hfill
%{\includegraphics[width=0.5\linewidth]{figure/new/mixture_male3}}\\
%\caption{{\color{red}Results of mixing 2 personalities on the faces of the European
%  male (left) and Middle-Eastern male (right).}}
%  \label{fig:mix2}
%\end{figure*}

%\begin{figure}[t]
 %\centering
 % \includegraphics[width=1\linewidth]{figure-new/3mix2}\\
  %\includegraphics[width=1\linewidth]{figure-new/3mix1}\\
  %\caption{Results of combining 3 personality impressions on the faces of the
   % European male (top) and African male (bottom). For each character,
    %faces synthesized with a single personality impression and combined
   % personality impressions are shown.}
%\label{fig:mix3} 
%\vspace{-6mm}
%\end{figure}

\subsection{Results and Discussion}
\label{sec:results}

\ssection{Different Faces.}  We test our approach to synthesizing
different faces to give different personality
impressions. Figure~\ref{fig:mass} shows two groups of the input faces and the
synthesized faces. For each input face, a face is synthesized using each of the $8$
impression types. Please refer to our supplementary material for more results of different races.

%The input faces consist of an \emph{European male}
%face, an \emph{African male} face, a \emph{Middle-Eastern male} face,
%an \emph{Asian female} face, and an \emph{European female} face.

We observe some interesting features that may result in the
corresponding personality impressions. For instance, comparing the
results of confident and unconfident faces, we observe that the
confident faces usually have a higher nose bridge and bigger eyes. In
addition, the eyebrows also look sharp and slightly slanted, which
make a person look like in a state of concentration. The mouth corners
lift slightly and the mouths show a subtle smile. As for the
unconfident faces, the eyebrows are generally dropping or furrowed,
showing a subtle sign of nervousness. The eye corners are also
dropping, and the eyes look tired. The cheeks generally look more
bonier. The mouths are also drooping, which could be perceived as signs
of frustration.

%\begin{figure*}[tbp]
 % \centering
%{\includegraphics[width=0.5\linewidth]{figure-new/mixture_male1}}%\hfill
%{\includegraphics[width=0.5\linewidth]{figure-new/mixture_male3}}
%\vspace{-5mm}
  %\caption{Results of combining 2 personality impression types on the %European
 % male (left) and Middle-Eastern male (right) faces.}
 % \label{fig:mix2}
%   \vspace{5mm}
%\end{figure*}

We observe that usually a combination of facial
features accounts for the personality impression of a face. As there
are as many as $160$ facial attributes, it is rather hard to manually
tune these attributes to model a face. The CNN classifiers effectively
learn the relationships between facial features and a personality
impression type, such that they can drive face synthesis by
personality impression automatically. 
%A human perceptual study on the
%synthesized faces is given in Section~\ref{sec:perceptual}.

\ssection{Multiple Personality Impressions.}  We also apply our approach to
synthesize faces with respect to multiple personality
impressions. Such faces could be useful in movies and games. For
example, it is common to have antagonists who look smart and
hostile. Our approach can be easily extended to synthesize such faces,
by optimizing a face with respect to multiple personality impression
costs, each of which corresponds to one personality impression type. Please refer to our supplementary material for details.

\begin{figure*}[t]
 \centering
 \vspace{-2mm}
  \includegraphics[width=\linewidth]{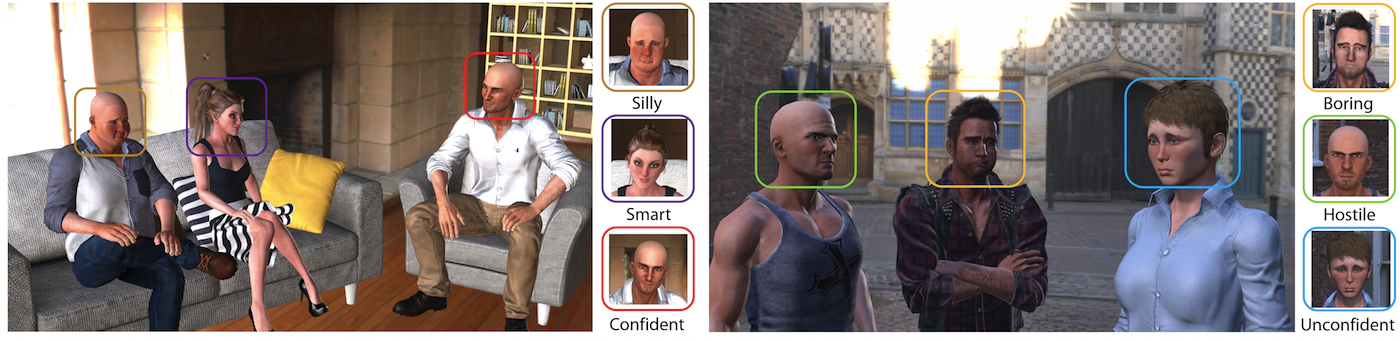}
\vspace{-3mm}
  \caption{Characters with faces synthesized by our approach. Left: a silly-looking
    man, a smart-looking lady and a confident-looking boss in an
    office. Right: a boring-looking man, a hostile-looking man and an
    unconfident-looking lady on a street.}
\label{fig:crowd}
\vspace{-4mm}
\end{figure*}

\ssection{Generating Crowds.}  Crowds of virtual characters showing
certain personality impressions are often needed in movies and
computer games. Our approach makes it very easy and convenient to
create such virtual characters. Figure~\ref{fig:crowd} shows two
examples of an office scene and a street scene showing virtual
character faces synthesized with personality impressions. Our
optimization approach could be employed for automatically synthesizing
virtual character faces (\eg, generating random hostile-looking
enemies in a 3D game) to enhance the realism of a virtual world.

\ssection{Remodeling 3D-reconstructed Faces.} Our approach can also be
used for remodeling 3D-reconstructed faces to give different types of
personality impression, which can help the user generate some personalized characters for 3D games and movies.
%Figure~\ref{fig:real} shows different faces
%created based on Obama and Putin. 
We reconstructed their 3D faces
based on their face images~\cite{blanz1999morphable}. We then applied
our approach to remodeling their faces with respect to different types
of personality impressions. In all cases, the similarity cost
constrains the synthesized faces to resemble the original faces of the
real persons. We show some examples in our supplementary material.

% The results show that
% our approach is effective on the synthesis of real people face models,
% which keeps similarity with the original face model and makes the
% model to show desired personalities by altering subtle facial
% features.
\vspace{-2mm}
\subsection{Perceptual Studies}
\label{sec:perceptual}
We conducted perceptual studies to evaluate the quality of our
results. The major goal is to verify if the perceived personality
impressions of the synthesized faces match with the personality
impression types. We recruited $160$ participants from different countries via Amazon
Turk. They are evenly distributed by gender and are aged $18$ to $50$. Each participant was shown some synthesized faces and was asked about the personality impression they perceived. Definitions of the personality
impression types from a dictionary were shown as reference. Our supplementary material contains all original data and the results of t-tests, discussed as follows.
%The perceptual studies include three parts
%discussed as follows. 

%\begin{figure}
 %\centering  
  %\includegraphics[width=1.0\linewidth]{figure-new/terser-3mixture4.png}
   % \vspace{-8mm}
  %\caption{Remodeling the 3D-reconstructed faces of Obama and Putin
   % with respect to different types of personality impressions.}
    %\vspace{-4mm}
  %\label{fig:real}
%\end{figure}

\noindent\textbf{Recognizing Face Personality Impression.}
\label{sec:recognizing}
 In this study, we want to verify if the personality impression types of the synthesized faces
agree with human impressions. We used the faces from
Figure~\ref{fig:mass}. Each of these faces was synthesized using a
single personality impression type and was voted by $40$ human
participants. In voting for the personality impression type of a face,
a participant needed to choose $1$ out of the $8$ personality
impression types used in our approach. In total, we obtained $1,600$
votes for $40$ faces.

\begin{figure}[t]
 \centering

  \includegraphics[height=4.5cm]{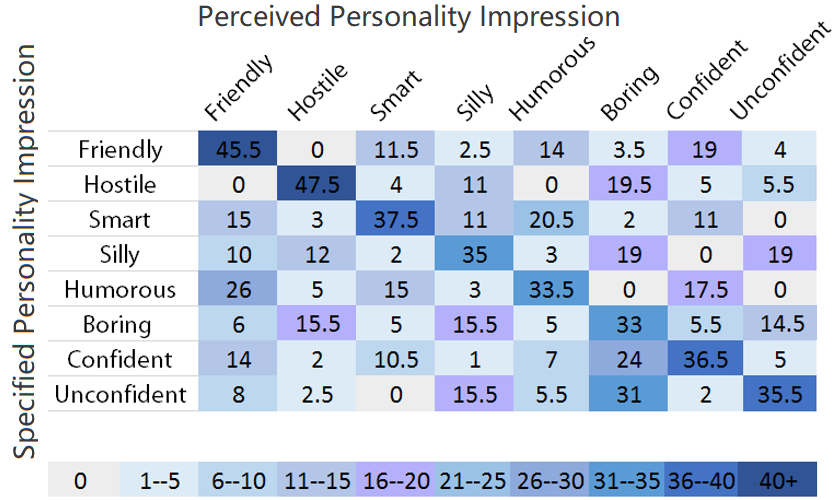}
    \vspace{-2mm}
  \caption{Accuracy of determining a single personality impression type of
    faces synthesized in the perceptual study. Percentages of votes are
    shown.}
     \vspace{-5mm}
\label{fig:Confusion-matrix1}
\end{figure}

Figure~\ref{fig:Confusion-matrix1} shows the results as a confusion
matrix. The average accuracy is about $38.0\%$ (compared to the
chance-level classification accuracy of $12.5\%$). For each
personality impression type, the matching type gets the highest number
of votes as shown by the diagonal. 

``Friendly'' and ``hostile'' receive a relatively high accuracy (about
$45\%-50\%$), probably because the facial features leading to such
personality impressions are more prominent and easily
recognizable. For example, participants usually perceive a face as
hostile-looking when they see dense moustache, slanted eyebrows and a
drooping mouth. For other personality impressions such as humorous and
boring, the accuracy is relatively lower (about $33\%)$), probably
because the facial features leading to such personality impressions
are less apparent, or because the participants do not have a strong
association between facial features and such personality
impressions. 

The facial features of some personalities are overlapped, which makes some people have several different, but similar personalities. For instance, a smart person may also look confident. Thus, participants may choose a similar which reduces the total accuracy.
%For example, it could be difficult to tell how a
%``boring'' person looks like just from one's face.

%The confusion matrix also visualizes the ambiguities. The major
%confusion exists between: friendly and confident, friendly and
%humorous, and hostile and boring. One possible reason is that the
%facial features leading to these personality impressions overlap
%partially. For example, friendly, confident and humorous-looking faces
%usually share the common feature of a lifting mouth corner, while
%hostile and boring-looking faces usually show a dropping mouth.

We also investigate how
human participants form the personality impressions of faces
synthesized with two personality impression types in our supplementary material.

%To conduct the study, we used $8$ faces synthesized by our approach, with each synthesized using a random combination of two personality impression types (\eg, friendly-confident, friendly-silly,
%hostile-confident), as shown in Figure~\ref{fig:mix2}. In the perceptual study, each face was voted by $40$
%participants, and a participant needed to choose two out of
%the eight personality impression types. In total, we obtained $320$
%data for the $8$ faces used.

%Figure~\ref{fig:Confusion-matrix2} shows the results. For most cases,
%the personality impression types used for synthesizing the faces take
%up the majority of the votes. Friendly-confident, friendly-smart and
%hostile-unconfident could be easily determined, whereas
%``friendly-silly'' and ``hostile-silly'' are harder to
%determine. Usually people are more confused among hostile, silly and
%boring, which is also observed in
%Figure~\ref{fig:Confusion-matrix1}. The reason might be that the
%interpretations of these personality impression types tend to overlap,
%so as the facial features associated with their impressions.

\noindent\textbf{Influence of Expression.}
\label{sec:influence}
Next we want to investigate whether facial expression changes will
affect the personality impression of the synthesized faces. For
example, does a face optimized to be hostile-looking still look
hostile with a happy smile? Such findings could bring interesting
insights for designing virtual character faces. We conducted an empirical study to investigate the effects of expressions on the synthesized faces. Please also refer to our supplementary material for details.

The average accuracy is about $33.4\%$ (a drop from $38\%$
on synthesized faces without any expression). Facial expressions do
have an impact on some personality impressions. For example, with an
angry expression, a face optimized to be friendly-looking may appear
hostile. The accuracy of the friendly (angry) face is $30.0\%$;
compared to the accuracy of $45.5\%$ on the friendly face without any
expression (Figure~\ref{fig:Confusion-matrix1}), the accuracy drops by
$15.5\%$. However, the personality impression on confident-looking faces seems
to be relatively unaffected by facial expressions. For instance, even
with an angry expression, a face optimized to look confident still has
$32.5\%$ votes for confident. This is probably because people have
strong associations between certain facial features and ``confident'', and those facial features are still apparent
under facial expression changes. 

Though this study is not comprehensive, it gives some good insights about the effects of expressions on personality impression. We believe
that a more comprehensive perceptual study will be an interesting
avenue for future research.

\section{Summary}
\label{sec:summary}
% In this paper, we have presented a novel approach to synthesize 3D
% faces driven by personality impression. Our approach optimizes the
% geometry and texture of the input face such that the synthesized face
% shows a desired personality impression specified by the user. During
% the optimization, our approach modifies the face iteratively by
% changing the control parameters of it. The resulting face is then
% evaluated by personality impression and similarity costs. The
% personality impression cost is formulated based on the prediction of a
% personality impression classifier, whereas the similarity cost is
% formulated based on a Siamese network to constrain the deformation of the
% synthesized face. Using our approach, the user can synthesize faces with regard to one or
% multiple personality impressions automatically, which could be useful
% for designing virtual characters for animations, games and virtual world
% applications. Our approach can also be employed to facilitate
% interactive face editing, where the designer modifies the main
% features of a 3D face in a few steps while our approach refines the
% fine-grained facial details based on the designer's modification, such that
% the resulting face shows a desired personality impression. Using our
% approach, designers can create faces with certain personality
% impressions in an easy and scalable manner.

\ssection{Limitations.} To stay focused on face's geometry and texture,
we do not consider the influence of hair, accessories or clothing
(\eg, hair style, hair color, hats, glasses) on personality
impression. Besides, speech and facial movements, as well as head and body
poses, can also influence the impression of one's personality, just as
experienced actors can change the personality impressions they make by
controlling speech, facial expression and body movements. While we
only focus on static facial features in this work, we refer the reader
to recent interesting efforts on adding personality to human motion~\cite{durupinar2017perform}.

\ssection{Future Work.} Our approach could be extended
to consider more personality impression types, other high-level
perceptual factors, or synthesizing faces of cartoon characters to give
certain personality impressions. Our approach follows the discriminative criteria to train the personality
impression classifier. For future work, it would be interesting to
investigate applying a deep generative network for synthesizing 3D
faces, as the adversarial training
approach (GAN)~\shortcite{goodfellow2014generative} has witnessed good
successes in 2D image generation.

\clearpage

%\renewcommand{\baselinestretch}{0.94}

%% The file named.bst is a bibliography style file for BibTeX 0.99c

\bibliographystyle{aaai}
\bibliography{aaai18}

\begin{thebibliography}{}

\bibitem[\protect\citeauthoryear{Blanz and Vetter}{1999}]{blanz1999morphable}
Blanz, V., and Vetter, T.
\newblock 1999.
\newblock A morphable model for the synthesis of 3d faces.
\newblock In {\em ACM SIGGRAPH},  187--194.
\newblock ACM Press/Addison-Wesley Publishing Co.

\bibitem[\protect\citeauthoryear{Blanz and Vetter}{2003}]{blanz2003face}
Blanz, V., and Vetter, T.
\newblock 2003.
\newblock Face recognition based on fitting a 3d morphable model.
\newblock {\em IEEE PAMI} 25(9):1063--1074.

\bibitem[\protect\citeauthoryear{Chopra, Hadsell, and
  LeCun}{2005}]{chopra2005learning}
Chopra, S.; Hadsell, R.; and LeCun, Y.
\newblock 2005.
\newblock Learning a similarity metric discriminatively, with application to
  face verification.
\newblock In {\em CVPR}.
\newblock IEEE.

\bibitem[\protect\citeauthoryear{Durupinar \bgroup et al\mbox.\egroup
  }{2017}]{durupinar2017perform}
Durupinar, F.; Kapadia, M.; Deutsch, S.; Neff, M.; and Badler, N.~I.
\newblock 2017.
\newblock Perform: Perceptual approach for adding ocean personality to human
  motion using laban movement analysis.
\newblock {\em TOG} 36(1):6.

\bibitem[\protect\citeauthoryear{Eisenthal, Dror, and
  Ruppin}{2006}]{eisenthal2006facial}
Eisenthal, Y.; Dror, G.; and Ruppin, E.
\newblock 2006.
\newblock Facial attractiveness: Beauty and the machine.
\newblock {\em Neural Computation} 18(1):119--142.

\bibitem[\protect\citeauthoryear{Goodfellow \bgroup et al\mbox.\egroup
  }{2014}]{goodfellow2014generative}
Goodfellow, I.; Pouget-Abadie, J.; Mirza, M.; Xu, B.; Warde-Farley, D.; Ozair,
  S.; Courville, A.; and Bengio, Y.
\newblock 2014.
\newblock Generative adversarial nets.
\newblock In {\em NIPS}. Curran Associates, Inc.
\newblock  2672--2680.

\bibitem[\protect\citeauthoryear{Gray \bgroup et al\mbox.\egroup
  }{2010}]{gray2010predicting}
Gray, D.; Yu, K.; Xu, W.; and Gong, Y.
\newblock 2010.
\newblock Predicting facial beauty without landmarks.
\newblock {\em ECCV}  434--447.

\bibitem[\protect\citeauthoryear{Hassin and Trope}{2000}]{hassin2000facing}
Hassin, R., and Trope, Y.
\newblock 2000.
\newblock Facing faces: studies on the cognitive aspects of physiognomy.
\newblock {\em Journal of Personality and Social Psychology} 78(5):837.

\bibitem[\protect\citeauthoryear{Hu \bgroup et al\mbox.\egroup
  }{2017}]{hu2017avatar}
Hu, L.; Saito, S.; Wei, L.; Nagano, K.; Seo, J.; Fursund, J.; Sadeghi, I.; Sun,
  C.; Chen, Y.-C.; and Li, H.
\newblock 2017.
\newblock Avatar digitization from a single image for real-time rendering.
\newblock {\em TOG} 36(6):195.

\bibitem[\protect\citeauthoryear{Huang \bgroup et al\mbox.\egroup
  }{2007}]{huang2007labeled}
Huang, G.~B.; Ramesh, M.; Berg, T.; and Learned-Miller, E.
\newblock 2007.
\newblock Labeled faces in the wild: A database for studying face recognition
  in unconstrained environments.
\newblock {\em Technical Report 07-49}.

\bibitem[\protect\citeauthoryear{Huang \bgroup et al\mbox.\egroup
  }{2017}]{huang2017shape}
Huang, H.; Kalogerakis, E.; Yumer, E.; and Mech, R.
\newblock 2017.
\newblock Shape synthesis from sketches via procedural models and convolutional
  networks.

\bibitem[\protect\citeauthoryear{Joo, Steen, and Zhu}{2015}]{joo2015automated}
Joo, J.; Steen, F.~F.; and Zhu, S.-C.
\newblock 2015.
\newblock Automated facial trait judgment and election outcome prediction:
  Social dimensions of face.
\newblock In {\em ICCV}.

\bibitem[\protect\citeauthoryear{Kalogerakis \bgroup et al\mbox.\egroup
  }{2012}]{kalogerakis2012probabilistic}
Kalogerakis, E.; Chaudhuri, S.; Koller, D.; and Koltun, V.
\newblock 2012.
\newblock A probabilistic model for component-based shape synthesis.
\newblock {\em TOG} 31(4):55.

\bibitem[\protect\citeauthoryear{Le, Why, and Ashraf}{2011}]{le2011shape}
Le, N.; Why, Y.; and Ashraf, G.
\newblock 2011.
\newblock Shape stylized face caricatures.
\newblock {\em Advances in Multimedia Modeling}  536--547.

\bibitem[\protect\citeauthoryear{Marsella \bgroup et al\mbox.\egroup
  }{2013}]{marsella2013virtual}
Marsella, S.; Xu, Y.; Lhommet, M.; Feng, A.; Scherer, S.; and Shapiro, A.
\newblock 2013.
\newblock Virtual character performance from speech.
\newblock In {\em Proceedings of Eurographics Symposium on Computer Animation},
   25--35.
\newblock ACM.

\bibitem[\protect\citeauthoryear{Mischel}{2013}]{mischel2013personality}
Mischel, W.
\newblock 2013.
\newblock {\em Personality and assessment}.
\newblock Psychology Press.

\bibitem[\protect\citeauthoryear{Naumann \bgroup et al\mbox.\egroup
  }{2009}]{naumann2009personality}
Naumann, L.~P.; Vazire, S.; Rentfrow, P.~J.; and Gosling, S.~D.
\newblock 2009.
\newblock Personality judgments based on physical appearance.
\newblock {\em Personality and Social Psychology Bulletin} 35(12):1661--1671.

\bibitem[\protect\citeauthoryear{Over and Cook}{2018}]{over2018spontaneous}
Over, H., and Cook, R.
\newblock 2018.
\newblock Where do spontaneous first impressions of faces come from?
\newblock {\em Cognition} 170:190--200.

\bibitem[\protect\citeauthoryear{Paysan \bgroup et al\mbox.\egroup
  }{2009}]{bfm09}
Paysan, P.; Knothe, R.; Amberg, B.; Romdhani, S.; and Vetter, T.
\newblock 2009.
\newblock A 3d face model for pose and illumination invariant face recognition.
\newblock In {\em Advanced video and signal based surveillance, 2009. AVSS'09.
  Sixth IEEE International Conference on},  296--301.
\newblock Ieee.

\bibitem[\protect\citeauthoryear{Ramamoorthi and
  Hanrahan}{2001}]{ramamoorthi2001efficient}
Ramamoorthi, R., and Hanrahan, P.
\newblock 2001.
\newblock An efficient representation for irradiance environment maps.
\newblock In {\em SIGGRAPH},  497--500.
\newblock ACM.

\bibitem[\protect\citeauthoryear{Ritchie \bgroup et al\mbox.\egroup
  }{2015}]{ritchie2015controlling}
Ritchie, D.; Mildenhall, B.; Goodman, N.~D.; and Hanrahan, P.
\newblock 2015.
\newblock Controlling procedural modeling programs with stochastically-ordered
  sequential monte carlo.
\newblock {\em TOG} 34(4):105.

\bibitem[\protect\citeauthoryear{Saito \bgroup et al\mbox.\egroup
  }{2017}]{saito2016photorealistic}
Saito, S.; Wei, L.; Hu, L.; Nagano, K.; and Li, H.
\newblock 2017.
\newblock Photorealistic facial texture inference using deep neural networks.
\newblock In {\em CVPR}.
\newblock IEEE.

\bibitem[\protect\citeauthoryear{Sohre \bgroup et al\mbox.\egroup
  }{2018}]{Nicholas2018PVL}
Sohre, N.; Adeagbo, M.; Helwig, N.; Lyford-Pike, S.; and Guy, S.
\newblock 2018.
\newblock Pvl: A framework for navigating the precision-variety trade-off in
  automated animation of smiles.
\newblock {\em AAAI}.

\bibitem[\protect\citeauthoryear{Suwajanakorn, Seitz, and
  Kemelmacher-Shlizerman}{2015}]{Suwajanakorn_2015_ICCV}
Suwajanakorn, S.; Seitz, S.~M.; and Kemelmacher-Shlizerman, I.
\newblock 2015.
\newblock What makes tom hanks look like tom hanks.
\newblock In {\em ICCV}.

\bibitem[\protect\citeauthoryear{Szegedy \bgroup et al\mbox.\egroup
  }{2015}]{szegedy2015going}
Szegedy, C.; Liu, W.; Jia, Y.; Sermanet, P.; Reed, S.; Anguelov, D.; Erhan, D.;
  Vanhoucke, V.; and Rabinovich, A.
\newblock 2015.
\newblock Going deeper with convolutions.
\newblock In {\em CVPR},  1--9.

\bibitem[\protect\citeauthoryear{Talton \bgroup et al\mbox.\egroup
  }{2011}]{talton2011metropolis}
Talton, J.~O.; Lou, Y.; Lesser, S.; Duke, J.; M{\v{e}}ch, R.; and Koltun, V.
\newblock 2011.
\newblock Metropolis procedural modeling.
\newblock {\em TOG} 30(2):11.

\bibitem[\protect\citeauthoryear{Tian and Xiao}{2016}]{tian2016facial}
Tian, L., and Xiao, S.
\newblock 2016.
\newblock Facial feature exaggeration according to social psychology of face
  perception.
\newblock {\em Computer Graphics Forum} 35(7):391--399.

\bibitem[\protect\citeauthoryear{Vernon \bgroup et al\mbox.\egroup
  }{2014}]{vernon2014modeling}
Vernon, R.~J.; Sutherland, C.~A.; Young, A.~W.; and Hartley, T.
\newblock 2014.
\newblock Modeling first impressions from highly variable facial images.
\newblock {\em Proceedings of the National Academy of Sciences}
  111(32):E3353--E3361.

\bibitem[\protect\citeauthoryear{Willis and Todorov}{2006}]{willis2006first}
Willis, J., and Todorov, A.
\newblock 2006.
\newblock First impressions making up your mind after a 100-ms exposure to a
  face.
\newblock {\em Psychological Science} 17(7):592--598.

\bibitem[\protect\citeauthoryear{Xu \bgroup et al\mbox.\egroup
  }{2015}]{xu2015new}
Xu, J.; Jin, L.; Liang, L.; Feng, Z.; and Xie, D.
\newblock 2015.
\newblock A new humanlike facial attractiveness predictor with cascaded
  fine-tuning deep learning model.
\newblock {\em arXiv:1511.02465}.

\bibitem[\protect\citeauthoryear{Yi \bgroup et al\mbox.\egroup
  }{2014}]{yi2014learning}
Yi, D.; Lei, Z.; Liao, S.; and Li, S.~Z.
\newblock 2014.
\newblock Learning face representation from scratch.
\newblock {\em arXiv:1411.7923}.

\bibitem[\protect\citeauthoryear{Zell \bgroup et al\mbox.\egroup
  }{2015}]{zell2015stylize}
Zell, E.; Aliaga, C.; Jarabo, A.; Zibrek, K.; Gutierrez, D.; McDonnell, R.; and
  Botsch, M.
\newblock 2015.
\newblock To stylize or not to stylize?: the effect of shape and material
  stylization on the perception of computer-generated faces.
\newblock {\em TOG} 34(6):184.

\bibitem[\protect\citeauthoryear{Zhu and Ramanan}{2012}]{zhu2012face}
Zhu, X., and Ramanan, D.
\newblock 2012.
\newblock Face detection, pose estimation, and landmark localization in the
  wild.
\newblock In {\em CVPR},  2879--2886.

\end{thebibliography}

\end{document}